\documentclass[aps,floatfix,superscriptaddress,showpacs,showkeys]{revtex4}

\usepackage{amssymb,amsmath,amsfonts,latexsym,graphicx,epsfig,fancybox}

\usepackage{color}
\usepackage{titlesec}
\titleformat{\section}{\large\bfseries}{\thesection}{1em}{}

\newcommand{\bea}{\begin{eqnarray}}
\newcommand{\ena}{\end{eqnarray}}
\newcommand{\nn}{\nonumber\\}

\newcommand{\be}{\begin{equation}}
\newcommand{\en}{\end{equation}}
\newcommand{\ed}{\end{document}}

\newcommand{\la}{\langle}
\newcommand{\ra}{\rangle}

\newcommand{\BR}{\mbox{$\cal{B}$}}

\begin{document}

\hfill MITP/18-047 (Mainz)

\title{Analyzing lepton flavor universality in the decays
\boldmath{
$\Lambda_b\to\Lambda_c^{(\ast)}(\frac12^\pm,\frac32^-) + \ell\,\bar\nu_\ell$} }

\author{Thomas Gutsche}
\affiliation{Institut f\"ur Theoretische Physik, Universit\"at T\"ubingen,\\
Kepler Center for Astro and Particle Physics,\\
Auf der Morgenstelle 14, D-72076, T\"ubingen, Germany}

\author{Mikhail A. Ivanov}
\affiliation{Bogoliubov Laboratory of Theoretical Physics, \\
Joint Institute for Nuclear Research, 141980 Dubna, Russia}

\author{J\"{u}rgen G. K\"{o}rner}
\affiliation{PRISMA Cluster of Excellence, Institut f\"{u}r Physik,
Johannes Gutenberg-Universit\"{a}t, \\
D-55099 Mainz, Germany}

\author{Valery~E.~Lyubovitskij}
\affiliation{Institut f\"ur Theoretische Physik, Universit\"at T\"ubingen,\\
Kepler Center for Astro and Particle Physics,\\
Auf der Morgenstelle 14, D-72076, T\"ubingen, Germany}
\affiliation{Departamento de F\'\i sica y Centro Cient\'\i fico
Tecnol\'ogico de Valpara\'\i so (CCTVal), Universidad T\'ecnica
Federico Santa Mar\'\i a, Casilla 110-V, Valpara\'\i so, Chile}
\affiliation{Department of Physics, Tomsk State University,
634050 Tomsk, Russia}

\author{Pietro Santorelli}
\affiliation{Dipartimento di Fisica ``E. Pancini'', Universit\`a di Napoli
Federico II, Complesso Universitario di Monte S. Angelo,
Via Cintia, Edificio 6, 80126 Napoli, Italy}
\affiliation{Istituto Nazionale di Fisica Nucleare, Sezione di
Napoli, 80126 Napoli, Italy}

\author{Chien-Thang Tran}
\affiliation{Dipartimento di Fisica ``E. Pancini'', Universit\`a di Napoli
Federico II, Complesso Universitario di Monte S. Angelo,
Via Cintia, Edificio 6, 80126 Napoli, Italy}
\affiliation{Istituto Nazionale di Fisica Nucleare, Sezione di
Napoli, 80126 Napoli, Italy}

\today

\begin{abstract}

 Lepton flavor universality can be tested in the semileptonic
 decays $\Lambda_b\to \Lambda_c^{(\ast)}$ where $\Lambda_c^{(\ast)}$
denotes either the ground state $\Lambda_c(2286)$ (with $J^P=1/2^+$)  
or its orbital excitations $\Lambda_c(2595)$ (with $J^P=1/2^-$) and
 $\Lambda_c(2625)$ (with $J^P=3/2^-$). 
We calculate the differential decay rates as well as the branching fractions
of these decays for both tauonic and muonic modes with form factors
obtained from a covariant confined quark model previously developed by us.
We present results for the rate ratios of the tauonic and muonic modes
which provide important tests of lepton flavor universality in
forthcoming experiments.

\end{abstract}


\maketitle

\section{Introduction}
\label{sec:intro}

In the standard model (SM) the three charged lepton generations 
($\ell=e,\mu,\tau$) together with their neutral and massless neutrinos
interact with the weak gauge bosons universally. This SM feature
is called lepton flavor universality. Recent experimental studies of the
leptonic $B\to \ell \nu_\ell$ and semileptonic $B \to D^{(*)}\ell \nu_\ell$ decays
have shown deviations from the predictions of lepton flavor universality in
the tauonic modes (for a review see Ref.~\cite{Ciezarek:2017yzh}).
If the observed deviations will be confirmed in future experiments, then it
will open a new window in the search for new physics (NP) beyond the SM. 
There are many theoretical papers which consider different
scenarios for the implementation of NP. Some studies  
extend the SM by introducing new particles and new interactions.
Other studies adopt a model-independent approach by adding a set
of NP operators to the effective Hamiltonian for the 
$b \to c \ell \nu_\ell$ transition. The numerical values of the new Wilson
coefficients are determined by fitting available experimental data.
Nonperturbative hadronic effects in the $B\to D^{(\ast)}$ transitions are
encoded in form factors which most often
are evaluated using methods of heavy quark effective theory
(HQET)~\cite{Neubert:1993mb,grozin2004heavy}. 

Lepton flavor universality can also be tested in the semileptonic 
$\Lambda_b\to \Lambda_c^{(\ast)}$ decays where $\Lambda_c^{(\ast)}$
denotes either the ground state $\Lambda_c(2286)$ (with $J^P=1/2^+$)  
or its orbital excitations $\Lambda_c(2595)$ (with $J^P=1/2^-$) and
$\Lambda_c(2625)$ (with $J^P=3/2^-$). Following previous work 
\cite{Leibovich:1997az} the authors of Ref.~\cite{Boer:2018vpx} discussed
the transition form factors for decays into the above two
excited $\Lambda_c^{(\ast)}$ states up to ${\cal O}(1/m_b,1/m_b)$ corrections
in the heavy quark mass expansion using methods of HQET. In their 
${\cal O}(1/m_b,1/m_b)$ analysis they showed
that all relevant form factors are expressable
through a single universal baryon Isgur-Wise function.

The semileptonic transition into the $J^P=1/2^+$ ground state
$\Lambda_c(2286)$ plus a heavy lepton pair $\tau\bar\nu_\tau$ has been studied
in a number of theoretical papers. Among these is Ref. \cite{DiSalvo:2018ngq}
the authors of which predicted the partial decay width starting from rather
general assumptions. The effects of five possible new physics interactions
were analyzed by adopting five different form factors.
In Ref. \cite{Li:2016pdv} the effects of adding  a single scalar or vector
leptoquark to the SM have been investigated. It was shown that the
best-fit solution for the Wilson coefficients
obtained in the corresponding $B$~decays leads to similar enhancements in
the branching fractions of the $\Lambda_b$~decays.
The decay widths as well as the ratios of branching fractions for the
$\tau$ and $e/\mu$ modes have been calculated
in Ref.~\cite{Azizi:2018axf} by using QCD sum rule form factors.
The decays $\Lambda_b \to p \ell^- \bar\nu_\ell$ and 
$\Lambda_b \to \Lambda_c \ell^- \bar\nu_\ell$ were studied in
Ref.~\cite{Detmold:2015aaa} by using form factors from lattice QCD with
relativistic heavy quarks. 
In Ref.~\cite{Shivashankara:2015cta} the authors presented predictions for
the ground-state to  ground-state~$\Lambda_b$~decay 
in extensions of the SM by adding NP operators with 
different Lorentz structures. 
The scope of Ref.~\cite{Shivashankara:2015cta} was extended in 
 Ref.~\cite{Datta:2017aue} by adding a tensor operator. Both
Refs.~\cite{Shivashankara:2015cta,Datta:2017aue} used form factors from
lattice QCD in their analysis. 
We mention that semileptonic  $\Lambda_b$ decays were also investigated in
the framework of a relativistic quark
model based on the quasipotential approach and a quark-diquark picture of
baryons~\cite{Faustov:2016pal}. 

%---------------------------------------------------------------------

In Ref.~\cite{Gutsche:2015mxa} we have provided a thorough analysis of the
decay  $\Lambda^0_b \to \Lambda^+_c(2286) + \tau^{-} +\bar \nu_{\tau}$
with particular emphasis on the lepton helicity flip and scalar contributions 
which vanish for zero lepton masses. We have calculated the total rate,
the differential decay distributions, the longitudinal, and transverse
polarizations of the daughter baryon $\Lambda^+_c$ as well as that of the
$\tau$~lepton, and the lepton-side forward-backward asymmetries. 

%---------------------------------------------------------------------

In a series of papers we have studied possible NP effects in the exclusive
decays $\bar{B}^0 \to D^{(\ast)} \tau^- \bar\nu_\tau$ 
and $B_c \to (J/\psi,\eta_c)\tau\nu_\tau$ including 
right-handed vector (axial), left- and right-handed
(pseudo)scalar, and tensor current
contributions~\cite{Tran:2018kuv,Ivanov:2017mrj,Ivanov:2016qtw}.
The $\bar{B}^0 \to D^{(\ast)}$  and $B_c \to (J/\psi,\eta_c)$ transition form
factors were calculated in the full kinematic $q^2$ range by employing
the covariant confined quark model (CCQM) previously developed by us. For more detail regarding the $R_{D^{(\ast)}}$ puzzle we refer to the recent study in Ref.~\cite{Feruglio:2018fxo} and references therein.

%---------------------------------------------------------------------

In Ref.~\cite{Gutsche:2017wag} we have calculated the invariant form
factors and the  helicity amplitudes for the transitions
$\Lambda_b~\to~\Lambda^{(\ast)}(J^P)~+~J/\psi$, where
the $\Lambda^{(\ast)}(J^P)$ are  $\Lambda(sud)$-type ground and excited states
with $J^P$ quantum numbers $J^P=1/2^\pm,3/2^\pm$. 

%---------------------------------------------------------------------

The purpose of the present paper is to calculate  the differential decay rates
and branching fractions of  the semileptonic
$\Lambda_b\to \Lambda_c^{(\ast)}$ decays in the SM for both the $\tau$ and
$\mu$ modes using form factors
evaluated in the  covariant confined quark model.

\section{Decay properties of the transitions
$\Lambda_b\to\Lambda_c^{(\ast)}(\frac12^\pm,\frac32^-) + \ell\,\bar\nu_\ell$ } 
\label{sec:definition}

The $J^P$~quantum numbers and the interpolating three-quark $(3q)$ currents
of the baryons involved in our calculations are shown in
Table~\ref{tab:baryons}. 
For the $P$-wave excitations with quantum numbers $J^P=1/2^-,3/2^-$
we have taken the simplest modifications of the ground state $J^P=1/2^+$
interpolating current.

\begin{table}[htb]
\begin{center}
\caption{Quantum numbers and interpolating currents
of charm and bottom baryons.}
\vspace*{.15cm}
\def\arraystretch{1.5}
\begin{tabular}{c|c|c|c}
\hline
\ \ Baryon \ \ & \ \ $J^P$ \ \ & \ \ Interpolating $3q$ current  \ \ 
& \ \ Mass (MeV) \ \ \\
\hline
$\Lambda_c(2286)$ & $\frac{1}{2}^+$ & $ \epsilon^{abc} \ c^a u^b C\gamma_5 d^c$ & 2286.46 \\
\hline
$\Lambda_c(2593)$ & $\frac{1}{2}^-$ & $ \epsilon^{abc} \ \gamma^5 c^a u^b C\gamma_5 d^c$  & 2592.25 \\
\hline
$\Lambda_c(2628)$ & $\frac{3}{2}^-$ & $ \epsilon^{abc} \ c^a u^b C\gamma_5\gamma_\mu d^c$ & 2628.11 \\
\hline
$\Lambda_b(5620)$ & $\frac{1}{2}^+$ & $ \epsilon^{abc} \ b^a u^b C\gamma_5 d^c$   & 5619.58 \\
\hline
\end{tabular}
\label{tab:baryons}
\end{center}
\end{table}

\begin{figure}[ht]
\begin{tabular}{lr}
  \includegraphics[width=0.40\linewidth]{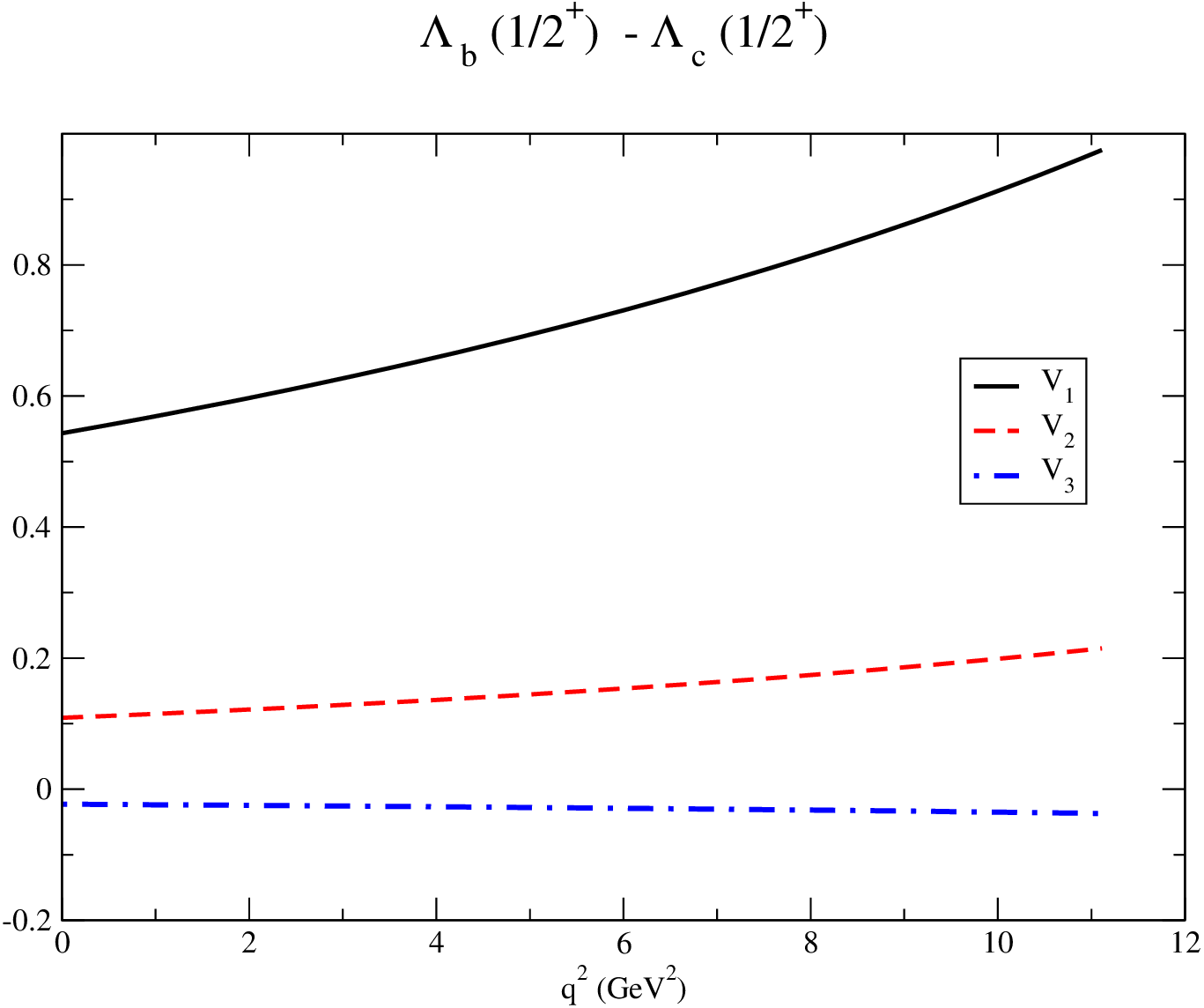}  \qquad\qquad &
  \includegraphics[width=0.40\linewidth]{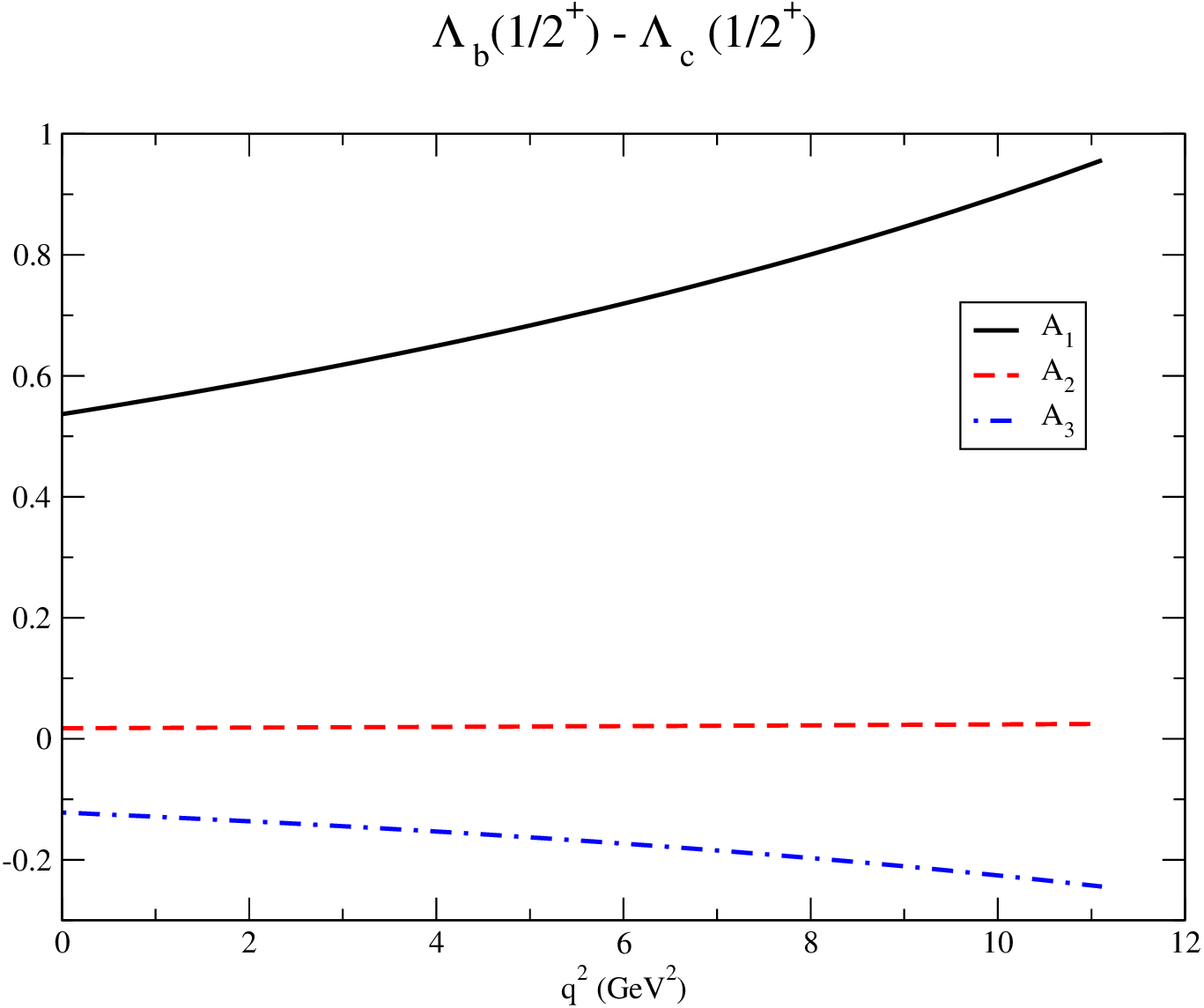}
\end{tabular}
\caption{Vector and axial form factors for the transition
$\Lambda_b(1/2^+)\to \Lambda_c(1/2^+)$.}
\label{fig:f12p}

\vspace*{.5cm}
\begin{tabular}{lr}
  \includegraphics[width=0.40\linewidth]{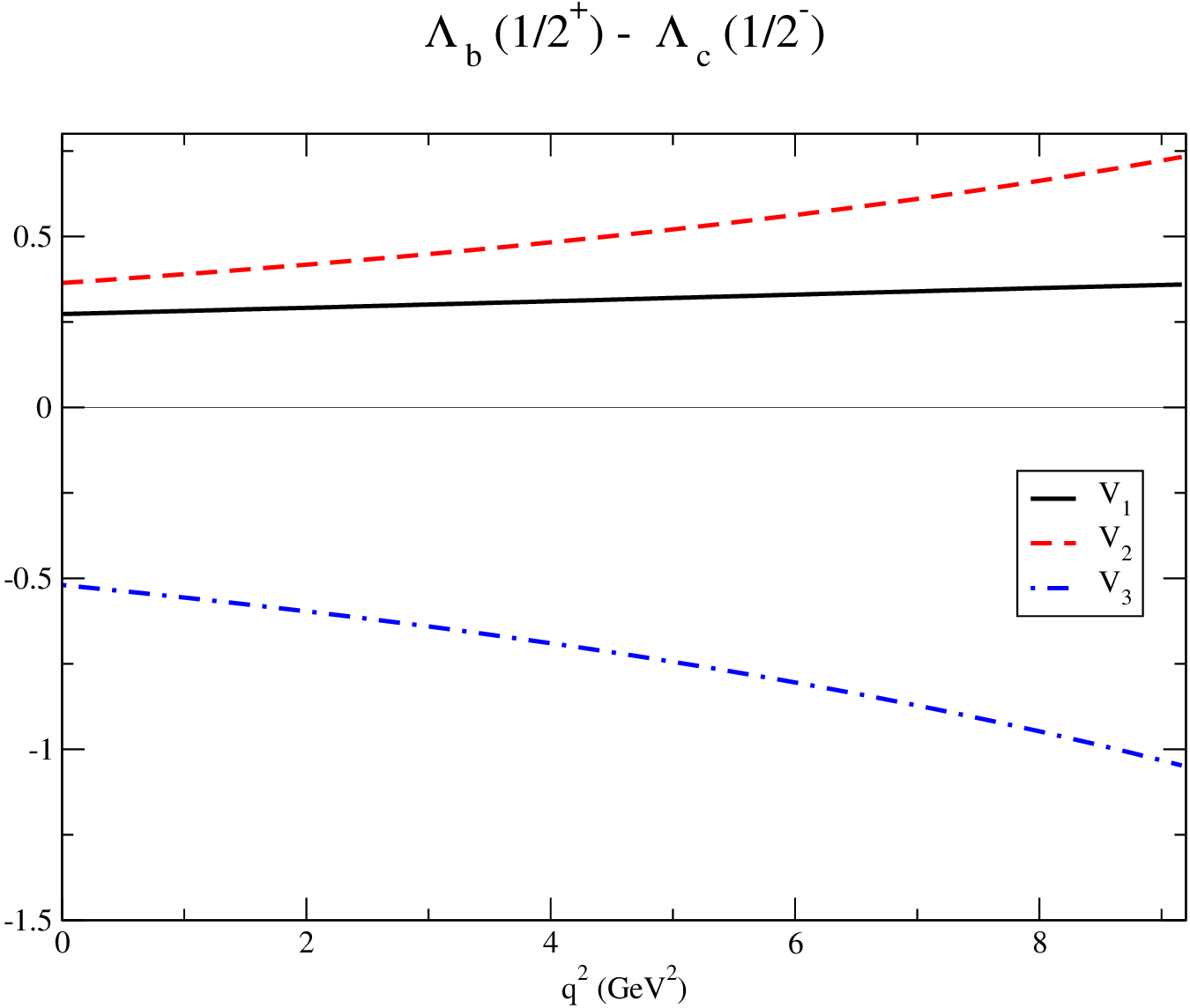} \qquad\qquad &
  \includegraphics[width=0.40\linewidth]{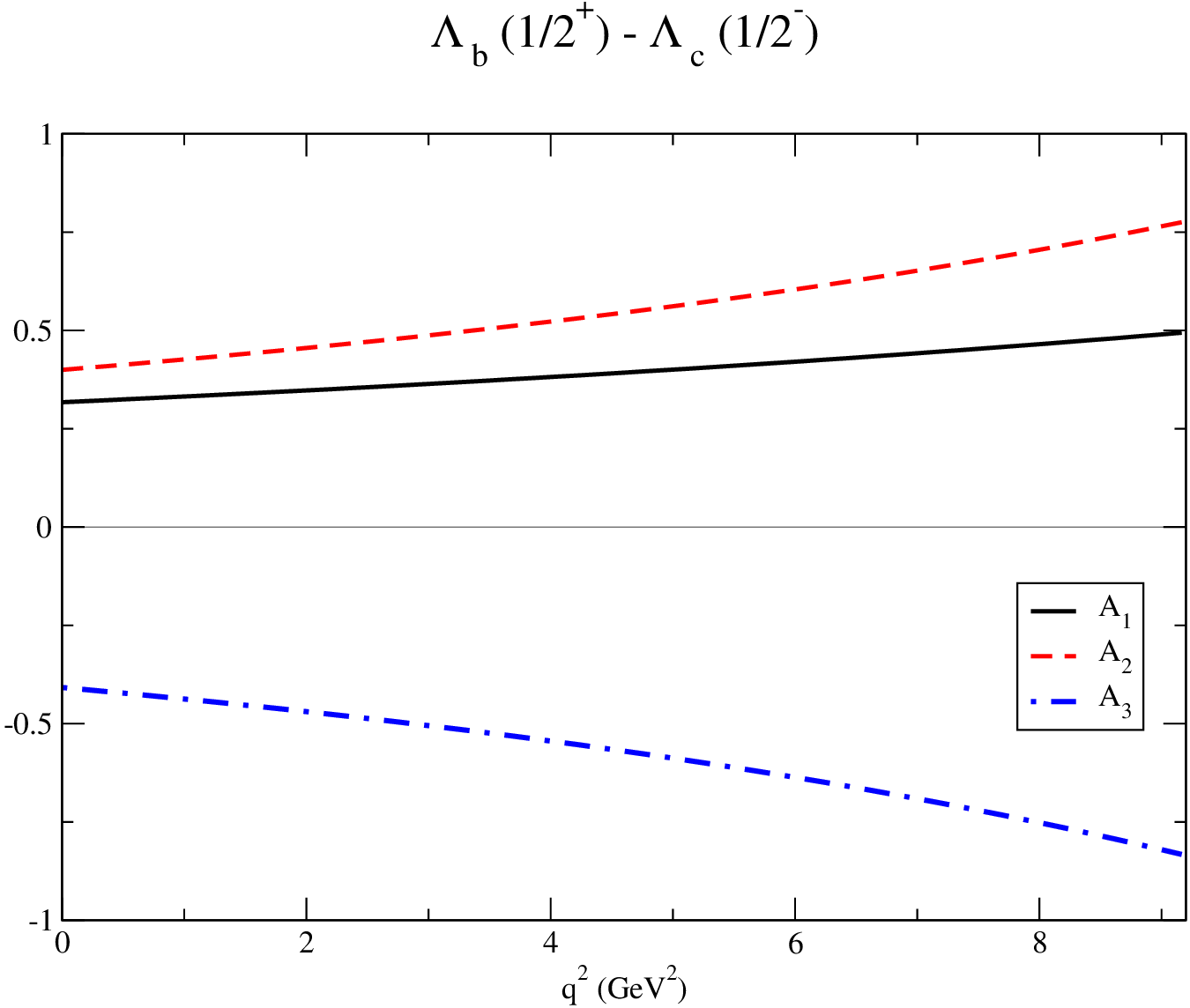}
\end{tabular}
\caption{Vector and axial form factors for the transition
$\Lambda_b(1/2^+)\to \Lambda_c(1/2^-)$.}
\label{fig:f12m}

\vspace*{.5cm}
\begin{tabular}{lr}
  \includegraphics[width=0.40\linewidth]{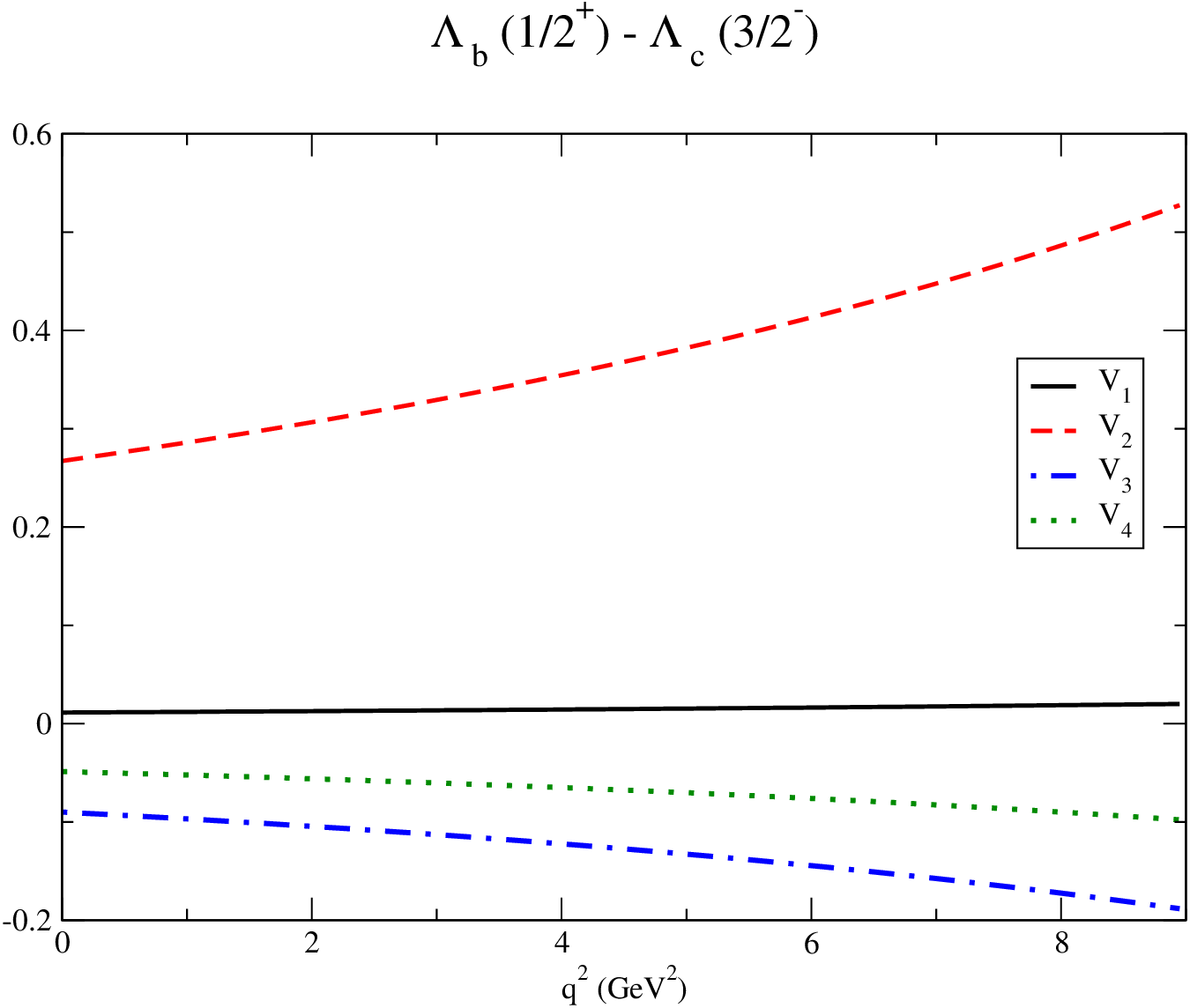}  \qquad\qquad &
  \includegraphics[width=0.40\linewidth]{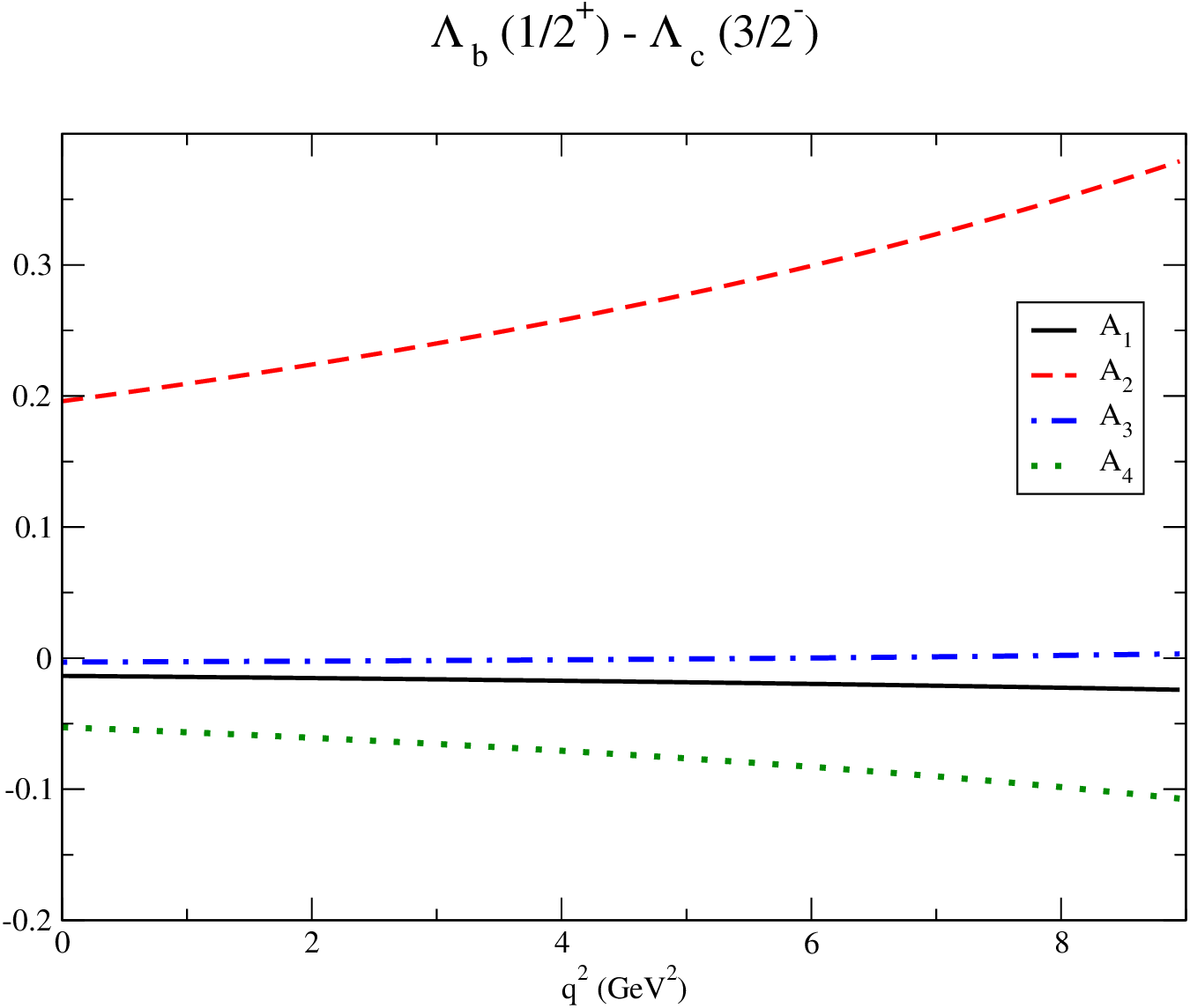}
\end{tabular}
\caption{Vector and axial form factors for the transition
$\Lambda_b(1/2^+)\to \Lambda_c(3/2^-)$.}
\label{fig:f32m}
\end{figure}

The hadronic matrix element $\la \Lambda_2 | \bar c O^\mu b | \Lambda_1 \ra$
$(O^\mu=\gamma^\mu(1-\gamma^5))$ is expressed in terms of six and eight
dimensionless invariant form factors
$F^{V/A}_i(q^2)$ for the transitions into the $\Lambda_2(1/2^\pm)$  
and $\Lambda_2(3/2^\pm)$ states, respectively. 
The details of their definition and their evaluation in the framework of
the CCQM can be found
in our paper~\cite{Gutsche:2017wag}. 
In Figs.~\ref{fig:f12p}-\ref{fig:f32m} we show the behavior of the
calculated form factors where we use a short-hand notation for the
form factors such that $V_i=F^V_i$ and $A_i=F^A_i$. 
We want to emphasize that our transition form factors are calculated using
finite quark masses. Thus they include the $1/m_b$- and $1/m_c$-corrections
considered in Ref.~\cite{Boer:2018vpx} as well as all higher powers of the
heavy mass expansion.

For the ground-state to ground-state transition $\Lambda_b \to \Lambda_c$
the finite mass form factors depicted in Fig.~\ref{fig:f12p}
show a close likeness to the limiting form factors of
the HQL (see, e.g., the review \cite{Kadeer:2005aq}).
In particular, the finite mass form factors $V_1(q^2)$ and $A_1(q^2)$ show
an approximate agreement with the zero
recoil normalization condition $V_1(q^2_{\rm max})~=~A_1(q^2_{\rm max})~=~1$ at zero
recoil $q^2_{\rm max}=(M_1-M_2)^2$. The form factors
$V_{2,3}(q^2)$ and $A_{2,3}(q^2)$ are predicted to be zero in the HQL.
Our finite mass form factors are small yet non zero.

%\graphicspath{{figures/}{eps/}}

The differential decay rate is given by 
(see Refs.~\cite{Gutsche:2015mxa,Gutsche:2017wag} for details)
\bea
\frac{d\Gamma}{dq^2} &=& \frac{G_F^2|V_{cb}|^2}{192 \pi^3}\,
\frac{(q^2-m_\ell^2)^2 |\mathbf{p_2}|}{M_1^2 q^2} \;{\cal H}_{\rm tot},
\label{eq:rate}\\[3ex]
{\cal H}_{\tfrac12\to\tfrac12}
&=&
|H_{\tfrac12 1}|^2 + |H_{-\tfrac12 -1}|^2 
+ |H_{\tfrac12 0}|^2 + |H_{-\tfrac12 0}|^2
\nn
&+&\frac{m^2_\ell}{2q^2}
\Big( 3\,|H_{\tfrac12 t}|^2 + 3\,|H_{-\tfrac12 t}|^2
     + |H_{\tfrac12 1}|^2 + |H_{-\tfrac12 -1}|^2 + |H_{\tfrac12 0}|^2 
+ |H_{-\tfrac12 0}|^2
\Big),
\label{eq:12}\\[2ex]
{\cal H}_{\tfrac12\to\tfrac32}
&=&
 |H_{\tfrac32 1}|^2 + |H_{-\tfrac32 -1}|^2 
+|H_{\tfrac12 1}|^2 + |H_{-\tfrac12 -1}|^2 
+ |H_{\tfrac12 0}|^2 + |H_{-\tfrac12 0}|^2
\nn
&+&\frac{m^2_\ell}{2q^2}
\Big( 3\,|H_{\tfrac12 t}|^2 + 3\,|H_{-\tfrac12 t}|^2
     + |H_{\tfrac32 1}|^2 + |H_{-\tfrac32 -1}|^2 
     + |H_{\tfrac12 1}|^2 + |H_{-\tfrac12 -1}|^2 
     + |H_{\tfrac12 0}|^2 + |H_{-\tfrac12 0}|^2
\Big)\,,
\ena
where $G_F = 1.16637 \times 10^{-5}$ GeV$^{-2}$ is the Fermi constant; 
$m_\ell$ is the charged lepton mass; 
$|\mathbf{p_2}|~=~\lambda^{1/2}(M_1^2,M_2^2,q^2)/(2 M_1)$ 
is the magnitude of the three-momentum of 
the daughter baryon in the rest frame of the parent baryon; 
$\lambda(x,y,z) = x^2 + y^2 + z^2 - 2xy - 2xz - 2yz$ is the 
kinematical triangle K\"allen function; 
$M_1$ and $M_2$ are the masses of the parent and daughter baryon,
respectively. The $H_{\lambda_1\,\lambda_W}$ denotes the helicity amplitudes which
are linearly related to the relativistic $\Lambda_b \to \Lambda_c^{(*)}$
transition form factors (for details see our recent
paper~\cite{Gutsche:2017wag}).    

\begin{figure}[ht]
\begin{tabular}{lr}
  \includegraphics[width=0.48\linewidth]{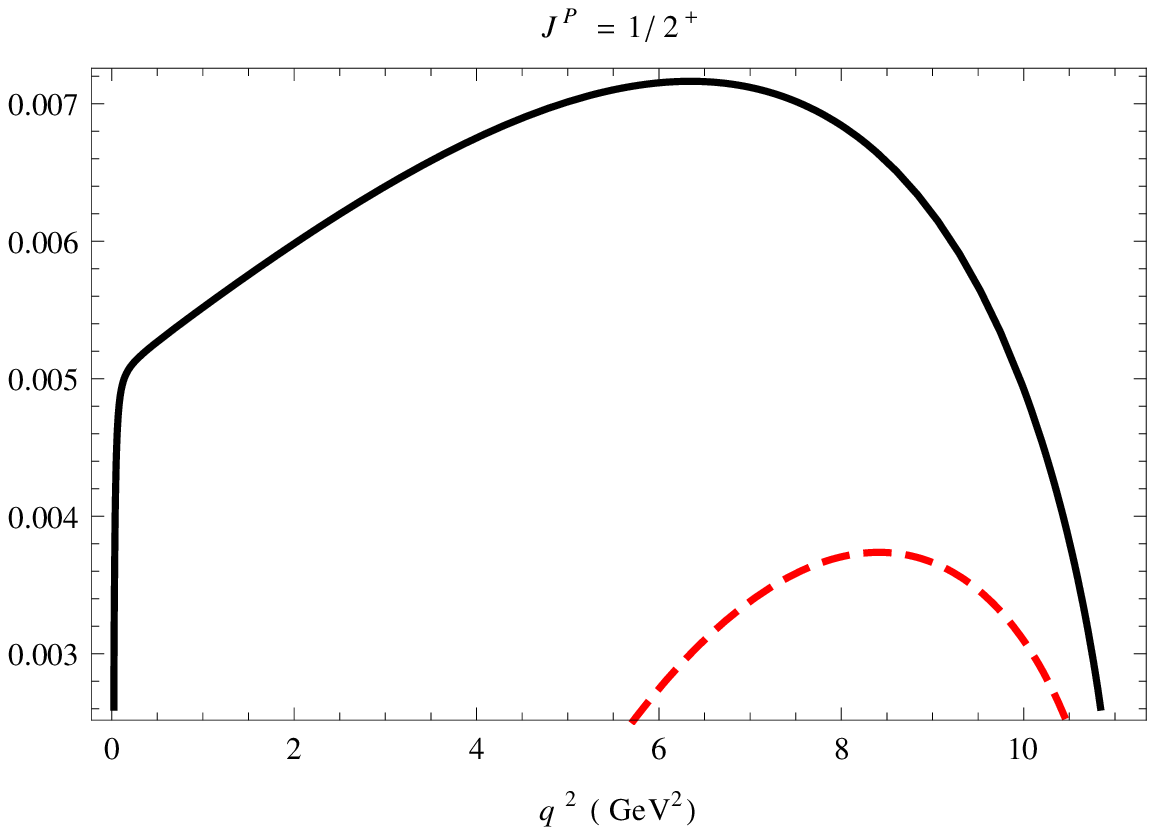} \quad &
  \includegraphics[width=0.48\linewidth]{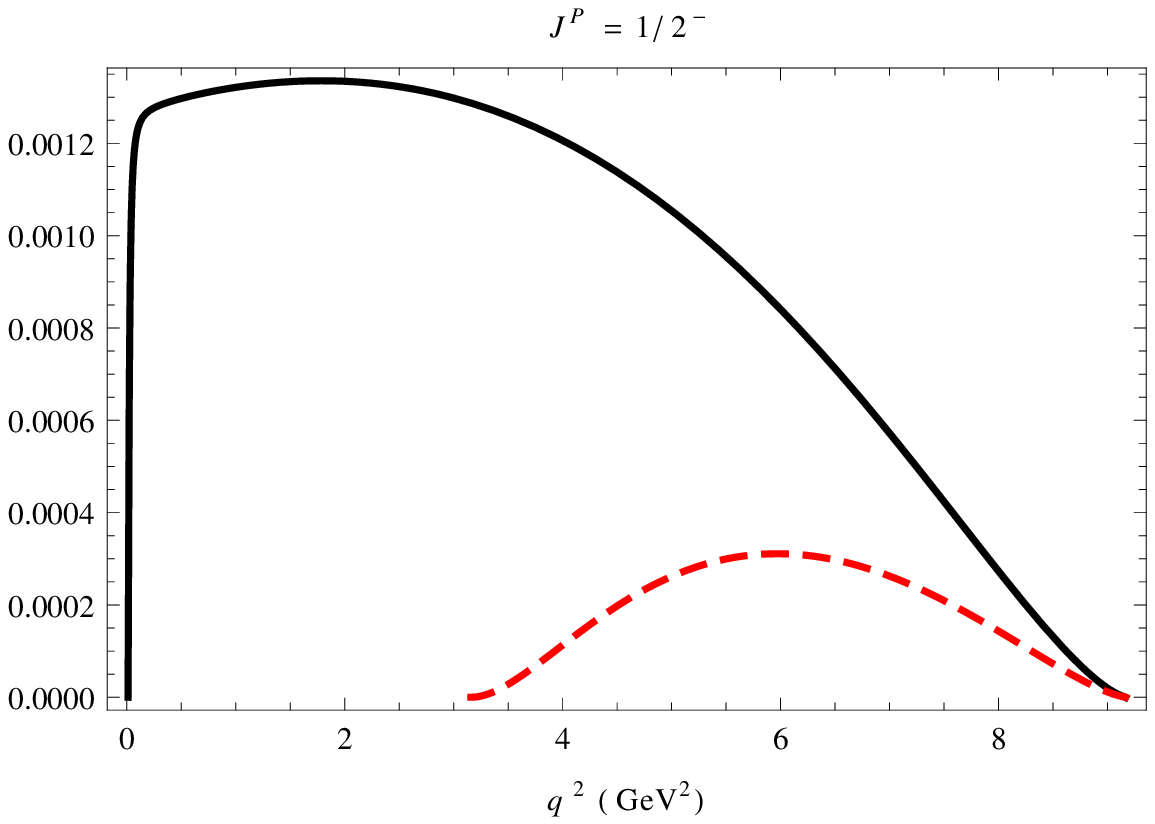} \\
  \includegraphics[width=0.48\linewidth]{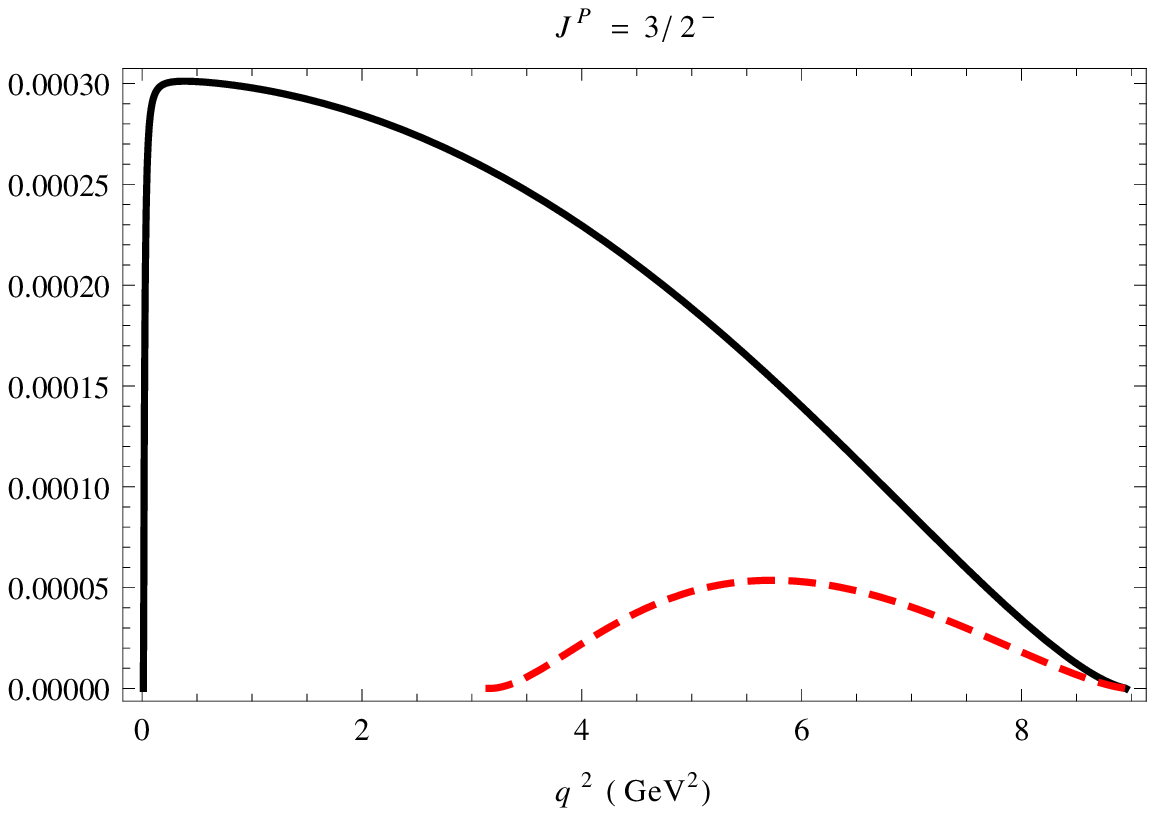} \quad & \\
\end{tabular}
\caption{The normalized differential decay rates for $\mu$ (solid) and
$\tau$ (dashed) modes.}
\label{fig:dBr}
\end{figure}

In Fig.~\ref{fig:dBr} we display the $q^2$~dependence of the normalized
differential decay rates 
in the full kinematical region for the $\mu$ and $\tau$ modes.
The $P$-wave factor $|\mathbf{p_2}|^3$ in the differential rate is clearly
visible for the $1/2^+ \to 1/2^-,3/2^-$ transitions at the zero recoil end
of the spectrum (see, e.g., the review \cite{Korner:1994nh}) .

In Table~\ref{tab:result}  we present our predictions for the semileptonic
branching fractions. We have used the central values for the lifetime and
the mass of the $\Lambda_b$  from
the Particle Data Group~\cite{Patrignani:2016xqp}
$\tau_{\Lambda_b}~=~(1.470~\pm~0.010)$~ps and
$M_{\Lambda_b}~=~(5619.58~\pm~0.17)$~MeV.
The value of the CKM matrix element is set to $|V_{cb}|~=~0.0405$.
We also display the numerical values for the ratio
\be
R(\Lambda_c^{(\ast)})
\equiv\frac{\BR(\Lambda^0_b \to \Lambda_c^{(\ast)\,+}\,\tau^-\bar\nu_\tau)}
           {\BR(\Lambda^0_b \to \Lambda_c^{(\ast)\,+}\,\mu^-\bar\nu_\mu)}. 
\label{eq:ratio}
\en
\begin{table}[ht] 
\begin{center}
\caption{The branching fractions (in $\%$) and the ratios 
$R(\Lambda_c^{(\ast)})$.} 
\label{tab:result}
\def\arraystretch{1.2}
\vspace*{0.2cm}
\begin{tabular}{c|ccc}
\hline
   & \qquad  $\Lambda_c^+(\tfrac12^+)$ \qquad & 
     \qquad  $\Lambda_c^{\ast\,+}(\tfrac12^-)$ \qquad & 
     \qquad $\Lambda_c^{\ast\,+}(\tfrac32^-)$ \\ 
\hline
e & \qquad 6.80 $\pm$ 1.36 \qquad & 
    \qquad 0.86 $\pm$ 0.17 \qquad & 
    \qquad 0.17 $\pm$ 0.03 \\
$\mu$ & \qquad 6.78 $\pm$ 1.36 \qquad & 
        \qquad 0.85 $\pm$ 0.17 \qquad & 
        \qquad 0.17 $\pm$ 0.03 \\
$\tau$ & \qquad 2.00 $\pm$ 0.40 \qquad & 
         \qquad 0.11 $\pm$ 0.02 \qquad & 
         \qquad 0.018 $\pm$ 0.004 \\
\hline
$R(\Lambda_c^{(\ast)})$ & \qquad 0.30 $\pm$ 0.06 \qquad & 
                       \qquad 0.13 $\pm$ 0.03 \qquad &  
                       \qquad 0.11 $\pm$ 0.02\\
\hline
\end{tabular}
\end{center}
\end{table}
\begin{table}[ht] 
\begin{center}
\caption{The ratio $R(\Lambda_c)$ calculated in various approaches.} 
\def\arraystretch{1.2}
\label{tab:ratio}
\vspace*{0.2cm}
\begin{tabular}{cccccccc}
\hline
  This work  & Ref.~\cite{DiSalvo:2018ngq} & Ref.~\cite{Li:2016pdv} 
 & Ref.~\cite{Azizi:2018axf} & Ref.~\cite{Detmold:2015aaa}  
 & Ref.~\cite{Shivashankara:2015cta} & Ref.~\cite{Faustov:2016pal} 
 & Ref.~\cite{Gutsche:2015mxa} \\
\hline
   0.30 $\pm$ 0.06 
&  \ [0.15,0.18] \ 
&  \ [0.27,0.33] \ 
&  \ 0.31 $\pm$ 0.11 \ 
&  \ 0.34 $\pm$ 0.01 \ 
&  \ 0.29 $\pm$ 0.02 \
&  \ 0.31 \ 
&  \ 0.29 \
\\
\hline
\end{tabular}
\end{center}
\end{table}

Finally, in Table~\ref{tab:ratio} we compare our result for the ratios
$R(\Lambda_c)$ with those obtained in other approaches.
Our finite mass result $R(\Lambda_c)~=~0.30~\pm~0.06$ is quite close to
the leading order HQL estimate $R(\Lambda_c)~=~r(x_\tau)/r(x_\mu)~=~0.244$
which follows from the results presented in Ref.~\cite{Kadeer:2005aq}.
The HQL rate ratio estimate $R(\Lambda_c)~=~r(x_\tau)/r(x_\mu)$
is true to ${\cal O}(\delta^2~=~0.178)$ where $\delta~=~(M_1-M_2)/(M_1+M_2)$.
The function $r(x_\ell)$ is given by
\be
r(x_\ell)=\sqrt{1-x_\ell^2}(1-\frac 92\, x_\ell^2-4x_\ell^4)-\frac{15}{2}x_\ell^4
\ln\frac{1-\sqrt{1-x_\ell^2}}{x_\ell}\,,
\en
where $x_\ell=m_\ell/(M_1-M_2)$.

In summary, we have calculated the ratios of the tauonic to the muonic modes
in the semileptonic decays of the bottom baryon~$\Lambda_b$ to the ground
state charm baryon~$\Lambda_c$ and the two lowest $P$-wave excitations with
$J^P=1/2^-,3/2^-$ quantum numbers. We are looking forward to the forthcoming
experimental results on the ratios of the $\tau$- and $\mu$-rates for the
three transitions that were analyzed in this paper. At a later stage one could
extend the comparison also to the differential $q^2$~distributions.

\begin{acknowledgments}

This work was funded  
by the German Bundesministerium f\"ur Bildung und Forschung (BMBF)
under Project 05P2015 - ALICE at High Rate (BMBF-FSP 202):
``Jet- and fragmentation processes at ALICE and the parton structure            
of nuclei and structure of heavy hadrons'',
by CONICYT (Chile) PIA/Basal FB0821, and by Tomsk State University
competitiveness improvement program under grant No. 8.1.07.2018.
M.A.I.\ acknowledges the support from  PRISMA cluster of excellence
(Mainz Uni.). M.A.I. and J.G.K. thank the Heisenberg-Landau Grant for
the partial support. P.S. acknowledges support by the Istituto Nazionale
di Fisica Nucleare, I.S. QFT$\_$HEP.

\end{acknowledgments}


\begin{thebibliography}{99}

%\cite{Ciezarek:2017yzh}
\bibitem{Ciezarek:2017yzh} 
  G.~Ciezarek, M.~Franco Sevilla, B.~Hamilton, R.~Kowalewski, 
  T.~Kuhr, V.~L\"uth and Y.~Sato,
  %``A Challenge to Lepton Universality in B Meson Decays,''
  Nature {\bf 546}, 227 (2017)
%  doi:10.1038/nature22346
  [arXiv:1703.01766 [hep-ex]].

\bibitem{Neubert:1993mb}
 M.~Neubert,
% Heavy quark symmetry,
 Phys.\ Rep.\ {\bf 245}, 259 (1994),
 arXiv:hep-ph/9306320. 

\bibitem{grozin2004heavy}
 A.~G.~Grozin,
 Heavy quark effective theory, 
 201 (Springer Science \& Business Media, 2014).
 

%-------------- Lb-Lc papers for discussion and comparison --------

 %\cite{Leibovich:1997az}
\bibitem{Leibovich:1997az}
  A.~K.~Leibovich and I.~W.~Stewart,
  %``Semileptonic Lambda(b) decay to excited Lambda(c) baryons
  %at order Lambda(QCD) / m(Q),''
  Phys.\ Rev.\ D {\bf 57} (1998) 5620
 % doi:10.1103/PhysRevD.57.5620
  [hep-ph/9711257].

%\cite{Boer:2018vpx}
\bibitem{Boer:2018vpx} 
  P.~B\"oer, M.~Bordone, E.~Graverini, P.~Owen, M.~Rotondo and D.~Van Dyk,
  %``Testing lepton flavour universality in semileptonic
  %$\Lambda_b \to \Lambda_c^*$ decays,''
  JHEP {\bf 1806}, 155 (2018)
%  doi:10.1007/JHEP06(2018)155
  [arXiv:1801.08367 [hep-ph]].
  
%\cite{DiSalvo:2018ngq}
\bibitem{DiSalvo:2018ngq} 
  E.~Di Salvo, F.~Fontanelli and Z.~J.~Ajaltouni,
  %``Detailed Study of the Decay $\Lambda_b\to\Lambda_c\tau\bar\nu_\tau$,''
  arXiv:1804.05592 [hep-ph].

%\cite{Li:2016pdv}
\bibitem{Li:2016pdv} 
  X.~Q.~Li, Y.~D.~Yang and X.~Zhang,
  %``$ {\varLambda}_b\to {\varLambda}_c\tau {\overline{\nu}}_{\tau } $ decay 
  % in scalar and vector leptoquark scenarios,''
  JHEP {\bf 1702}, 068 (2017)
%  doi:10.1007/JHEP02(2017)068
  [arXiv:1611.01635 [hep-ph]].

%\cite{Azizi:2018axf}
\bibitem{Azizi:2018axf} 
  K.~Azizi and J.~Y.~S\"ung\"u,
  %``Semileptonic $\Lambda_{b}\rightarrow \Lambda_{c}{\ell}\bar\nu_{\ell}$ 
  % Transition in Full QCD,''
  Phys.\ Rev.\ D {\bf 97}, 074007 (2018)
%  doi:10.1103/PhysRevD.97.074007
  [arXiv:1803.02085 [hep-ph]].

%\cite{Detmold:2015aaa}
\bibitem{Detmold:2015aaa} 
  W.~Detmold, C.~Lehner and S.~Meinel,
  %``$\Lambda_b \to p \ell^- \bar{\nu}_\ell$ and 
  % $\Lambda_b \to \Lambda_c \ell^- \bar{\nu}_\ell$ form factors 
  % from lattice QCD with relativistic heavy quarks,''
  Phys.\ Rev.\ D {\bf 92}, 034503 (2015)
%  doi:10.1103/PhysRevD.92.034503
  [arXiv:1503.01421 [hep-lat]].


%\cite{Shivashankara:2015cta}
\bibitem{Shivashankara:2015cta} 
  S.~Shivashankara, W.~Wu and A.~Datta,
  %``$\Lambda_b \to \Lambda_c \tau \bar{\nu}_{\tau}$ Decay in the Standard Model and with New Physics,''
  Phys.\ Rev.\ D {\bf 91}, 115003 (2015)
%  doi:10.1103/PhysRevD.91.115003
  [arXiv:1502.07230 [hep-ph]].

%\cite{Datta:2017aue}
\bibitem{Datta:2017aue} 
  A.~Datta, S.~Kamali, S.~Meinel and A.~Rashed,
  %``Phenomenology of ${\Lambda}_b\to {\Lambda}_c\tau {\overline{\nu}}_{\tau }$
  %  using lattice QCD calculations,''
  JHEP {\bf 1708}, 131 (2017)
%  doi:10.1007/JHEP08(2017)131
  [arXiv:1702.02243 [hep-ph]].

%\cite{Faustov:2016pal}
\bibitem{Faustov:2016pal} 
  R.~N.~Faustov and V.~O.~Galkin,
  %``Semileptonic decays of $\Lambda_b$ baryons in 
  % the relativistic quark model,''
  Phys.\ Rev.\ D {\bf 94}, 073008 (2016)
%  doi:10.1103/PhysRevD.94.073008
  [arXiv:1609.00199 [hep-ph]].


%\cite{Gutsche:2015mxa}
\bibitem{Gutsche:2015mxa} 
  T.~Gutsche, M.~A.~Ivanov, J.~G.~K\"orner, V.~E.~Lyubovitskij, 
  P.~Santorelli and N.~Habyl,
  %``Semileptonic decay $\Lambda_b \to \Lambda_c + \tau^- + \bar{\nu_\tau}$ 
  % in the covariant confined quark model,''
  Phys.\ Rev.\ D {\bf 91}, 074001 (2015)
  Erratum: [Phys.\ Rev.\ D {\bf 91}, 119907 (2015)]
%  doi:10.1103/PhysRevD.91.074001, 10.1103/PhysRevD.91.119907
  [arXiv:1502.04864 [hep-ph]].

%------------- our papers on New Physics -------------------------------

%\cite{Tran:2018kuv}
\bibitem{Tran:2018kuv}
  C.~T.~Tran, M.~A.~Ivanov, J.~G.~K\"orner and P.~Santorelli,
  %``Implications of new physics in the decays 
  % $B_c \to (J/\psi,\eta_c)\tau\nu$,''
  Phys.\ Rev.\ D {\bf 97}, 054014 (2018) 
%  doi:10.1103/PhysRevD.97.054014
  [arXiv:1801.06927 [hep-ph]].
  
%\cite{Ivanov:2017mrj}
\bibitem{Ivanov:2017mrj} 
  M.~A.~Ivanov, J.~G.~K\"orner and C.~T.~Tran,
  %``Probing new physics in $\bar{B}^0 \to D^{(\ast)} \tau^- \bar\nu_{\tau}$ 
  % using the longitudinal, transverse, and normal polarization components 
  % of the tau lepton,''
  Phys.\ Rev.\ D {\bf 95}, 036021 (2017)
%  doi:10.1103/PhysRevD.95.036021
  [arXiv:1701.02937 [hep-ph]].

%\cite{Ivanov:2016qtw}
\bibitem{Ivanov:2016qtw} 
  M.~A.~Ivanov, J.~G.~K\"orner and C.~T.~Tran,
  %``Analyzing new physics in the decays 
  % $\bar{B}^0 \to D^{(\ast)}\tau^-\bar\nu_{\tau}$ with form factors 
  % obtained from the covariant quark model,''
  Phys.\ Rev.\ D {\bf 94},  094028 (2016)
%  doi:10.1103/PhysRevD.94.094028
  [arXiv:1607.02932 [hep-ph]].

\bibitem{Feruglio:2018fxo} 
  F.~Feruglio, P.~Paradisi and O.~Sumensari,
  %``Implications of scalar and tensor explanations of $R_{D^{(\ast)}}$,''
  arXiv:1806.10155 [hep-ph].

  %------ $\Lambda_b \to \Lambda^{(\ast)}(\frac12^\pm,\frac32^\pm) + J/\psi$

  %\cite{Gutsche:2017wag}
\bibitem{Gutsche:2017wag} 
  T.~Gutsche, M.~A.~Ivanov, J.~G.~K\"orner, V.~E.~Lyubovitskij, 
  V.~V.~Lyubushkin and P.~Santorelli,
  %``Theoretical description of the decays 
  % $\Lambda_b \to \Lambda^{(\ast)}(\frac12^\pm,\frac32^\pm) + J/\psi$,''
  Phys.\ Rev.\ D {\bf 96}, 013003 (2017)
%  doi:10.1103/PhysRevD.96.013003
  [arXiv:1705.07299 [hep-ph]].

  %\cite{Kadeer:2005aq}
\bibitem{Kadeer:2005aq}
  A.~Kadeer, J.~G.~K\"orner and U.~Moosbrugger,
  %``Helicity analysis of semileptonic hyperon decays including
  % lepton mass effects,''
  Eur.\ Phys.\ J.\ C {\bf 59}  27 (2009)
%  doi:10.1140/epjc/s10052-008-0801-5
  [hep-ph/0511019].
  
  %\cite{Korner:1994nh}
\bibitem{Korner:1994nh}
  J.~G.~K\"orner, M.~Kr\"amer and D.~Pirjol,
  %``Heavy baryons,''
  Prog.\ Part.\ Nucl.\ Phys.\  {\bf 33} 787  (1994)
%  doi:10.1016/0146-6410(94)90053-1
  [hep-ph/9406359].
  
  %\cite{Patrignani:2016xqp}
\bibitem{Patrignani:2016xqp} 
  C.~Patrignani {\it et al.} [Particle Data Group],
  %``Review of Particle Physics,''
  Chin.\ Phys.\ C {\bf 40}, 100001 (2016).
%  doi:10.1088/1674-1137/40/10/100001

\end{thebibliography}
\end{document}